\documentclass[usenatbib]{mn2e}
\usepackage{graphicx}

\begin{document}
\topmargin-1cm

\newcommand\approxgt{\mbox{$^{>}\hspace{-0.24cm}_{\sim}$}}
\newcommand\approxlt{\mbox{$^{<}\hspace{-0.24cm}_{\sim}$}}
\newcommand{\be}{\begin{equation}}
\newcommand{\ee}{\end{equation}}
\newcommand{\bea}{\begin{eqnarray}}
\newcommand{\eea}{\end{eqnarray}}
\newcommand{\lexp}{\mathop{\langle}}
\newcommand{\rexp}{\mathop{\rangle}}
\newcommand{\rexpc}{\mathop{\rangle_c}}
\newcommand{\nbar}{\bar{n}}
\newcommand{\zmax}{Z_{\rm max}}
\newcommand{\amass}{10^{13} \,M_\odot}
\newcommand{\hmass}{10^{14} \,M_\odot}
\newcommand{\mmin}{M_{\rm{min}}}

\newcommand{\apj}{ApJ}
\newcommand{\apjl}{ApJL}
\newcommand{\apjs}{ApJS}
\newcommand{\mnras}{MNRAS}
\newcommand{\aap}{AAP}
\newcommand{\prd}{PRD}
\newcommand{\aj}{AJ}

\title{Constraints on the Merging Timescale of Luminous Red Galaxies,
  Or, Where Do All the Halos Go?}

\author[Conroy et al.]
{Charlie Conroy${}^1$\thanks{cconroy@astro.princeton.edu},
Shirley Ho${}^1$ and Martin White$^{2}$\\
$^{1}$ Department of Astrophysical Sciences, Peyton Hall,
Princeton University, Princeton, NJ 08544, USA.\\
$^{2}$ Departments of Physics and Astronomy, 601 Campbell Hall,
University of California Berkeley, CA 94720, USA.}
\date{\today}
\maketitle

\begin{abstract}
  In the $\Lambda$CDM cosmology dark matter halos grow primarily
  through the accretion of smaller halos.  Much of the mass in a halo
  of $\hmass$ comes in through accretion of $\sim\amass$ halos.  If
  each such halo hosted one luminous red galaxy (LRG) then the
  accretion of so many halos is at odds with the observed number of
  LRGs in clusters unless these accreted LRGs merge or disrupt on
  relatively short timescales ($\sim2$ Gyr).  These timescales are
  consistent with classical dynamical friction arguments, and imply
  that $2-3$ LRGs have merged or disrupted within each halo more
  massive than $\hmass$ by $z=0$.  The total amount of stellar mass
  brought into these massive halos by $z=0$ is consistent with
  observations once the intracluster light (ICL) is included.  If
  disrupted LRGs build up the ICL, then the hierarchical growth of
  massive halos implies that a substantial amount of ICL should also
  surround satellite LRGs, as suggested by recent observations of the
  Virgo cluster.  Finally, we point out that these results are
  entirely consistent with a non-evolving clustering strength and halo
  occupation distribution, and note that observations of the latter in
  fact support the hypothesis that merging/disruption of massive
  galaxies does indeed take place at late times.

\end{abstract}


\section{Introduction} \label{sec:introduction}

The formation and evolution of massive red galaxies provide a critical
testing ground for modern theories of galaxy formation based on
hierarchical merging of dark matter halos.  Ongoing growth of massive
halos via mergers is a generic feature of hierarchical models, such as
cold dark matter (CDM).  However evidence for the ongoing assembly of
massive galaxies is at best inconclusive.  Evolution in the galaxy
stellar mass and luminosity functions at the massive/luminous end
appears quite modest since $z=1$ \citep[e.g.][]{Drory04, Bundy05,
  Borch06, Fontana06,Faber06, Willmer06, Brown07, Caputi06, Wake06}
though estimates of the merger rate of massive galaxies present a less
consistent picture \citep{vanDokkum05,Bell06,Masjedi06, MWhite07}.  If,
as theory predicts, massive halos are constantly accreting halos that
are themselves hosts of massive galaxies, what is the fate of these
accreted galaxies?

Two physical effects can cause satellite galaxies to `disappear' from
an observational sample.  Tidal forces acting on a satellite as it
orbits in the host halo potential can cause it to \emph{disrupt}.  At
the same time, dynamical friction (DF) causes a satellite to lose
energy to the background dark matter halo and eventually causes the
satellite to sink toward the center and \emph{merge} with the central
galaxy of the host halo.  While such notions, and their relevance to
the evolution of galaxies within clusters, have been known for decades
\citep[e.g.][]{Chandrasekhar43, Ostriker75, Ostriker77, Merritt84},
accurate merger times of satellite galaxies have been historically
hard to calculate, and are poorly constrained observationally.
Unfortunately, the problem cannot at present by circumvented by
brute-force simulations due both to severe resolution requirements and
the uncertain effects of baryon condensation on the survival of
satellite halos \citep[see e.g.][]{Moore99, Klypin99, Diemand04,
  Gao04a, Gao04b, Reed05}.  In addition, while DF is usually
considered in a collisionless medium (such as dark matter), DF acting
in a collisional medium (such as intracluster gas) is stronger
(weaker) than in the collisionless case for satellites traveling at
supersonic (subsonic) speeds \citep{Ostriker99}.  Observational
constraints on the merging timescale of satellites would hence provide
valuable insight into this complex dynamical process.

This paper explores observational constraints on the average merging
timescale of luminous red galaxies (LRGs).  We assign LRGs to dark
matter halos that have grown more massive than $M\sim\amass$ and use
an $N$-body simulation to follow their accretion onto larger dark
matter halos with $z=0$ mass comparable to observed rich groups and
clusters $(M>\hmass)$.  Comparison with the observed multiplicity
function of LRGs at $z\sim0.3$ implies that accreted LRGs must merge
on timescales comparable to those predicted by Chandrasekhar's formula
($\sim2$ Gyr).  While this may not be surprising, the relative flood
of massive halos onto more massive halos implies that a substantial
number of LRGs have disrupted over the history of the Universe.
Though these numbers may at first glance appear large (on average
$2-3$ disrupted LRGs per $z=0$ halo with $M>\hmass$) we show that the
total stellar mass brought in by these accreted LRGs is consistent
with the observed stellar mass in clusters so long as one counts both
observed massive galaxies and the observed intracluster light.

The following sections describe in more detail the salient accretion
properties of massive dark matter halos ($\S$\ref{s:accretion}), the
inferred merging timescale of LRGs, if LRGs correspond to massive
halos ($\S$\ref{s:merge}), and the implied total stellar mass brought
into massive $z\sim0$ dark matter halos by these accreted LRGs
($\S$\ref{s:stellar}).  We conclude in $\S$\ref{s:conc}.  Throughout
we assume a flat $\Lambda$CDM cosmology with $(\Omega_m,
\Omega_{\Lambda}, h,\sigma_8) = (0.25,0.75,0.72,0.8)$, and use a
virial mass definition, $M_{200}$, corresponding to the mass contained
within a region that has mean density equal to $200\times$ the
critical density \citep[see e.g.][]{Evrard07}.

\section{The Accretion History of Massive Dark Matter
  Halos}\label{s:accretion}

\begin{figure}
\begin{center}
\resizebox{3.5in}{!}{\includegraphics{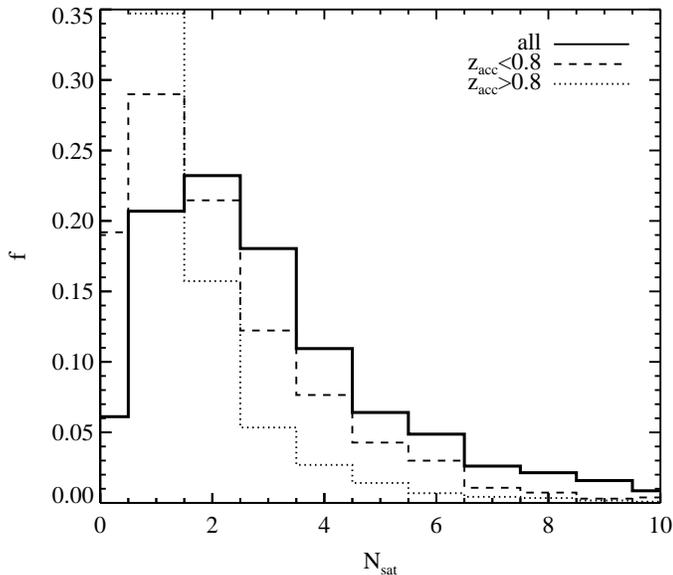}}
\end{center}
\caption{Fraction of halos at $z=0$ with $M>\hmass$, $f$, that have
  absorbed $N_{\rm sat}$ halos with mass $>\amass$.  The thick solid line
  is for halos accreted at all epochs, while the dashed (dotted) line
  indicates only those halos accreted after (before) $z=0.8$,
  i.e. when the Universe was half its present age for our assumed
  cosmology.  No distinction is made here between halos that have
  dissolved and halos that remain as bound satellites.  On average,
  $3.2$ halos with mass $>\amass$ have accreted onto these more
  massive halos by $z=0$.}
\label{fig:mergers}
\end{figure}

A robust expectation of a Universe dominated by cold dark matter is
the hierarchical growth of structure, and in particular the growth of
dark matter halos via the accumulation of smaller halos.  An
illustrative example of the accretion history of dark matter halos is
shown in Figure \ref{fig:mergers}.  There we plot the multiplicity
function for halos more massive than $10^{14}M_\odot$, i.e.~the
distribution of the number of halos with $M>\amass$ that have accreted
onto halos with $z=0$ mass greater than $10^{14} M_\odot$ (see the
appendix for details regarding the simulation used to compile this
information).  We refer to these more massive halos as ``hosts''
throughout.  There are 2339 such hosts in our simulation at $z=0$,
corresponding to a number density of $\sim 2 \times 10^{-5}$
Mpc$^{-3}$.  Note that these distributions are not symmetric.  On
average, halos more massive than $\hmass$ have been bombarded by $3.2$
halos with mass $>\amass$ over a Hubble time.  Here we do not
distinguish between halos that were accreted directly onto the host
halo and those that were accreted onto an intermediate halo that later
accreted onto the host, although such a distinction will be utilized
in the following sections.  The accretion of such massive halos is
roughly equally important both at low and high redshift: on average
two such halos have been accreted at $z<0.8$ (the Universe was about
half its present age at $z=0.8$ for the cosmology assumed herein).

Halos more massive than $\sim\amass$ are expected to contain at least
one massive galaxy at their center \citep{Zehavi05} even at moderate
redshifts \citep[e.g.][]{Yan03,Yan04,Coil06b}.  From Figure
\ref{fig:mergers} we are lead to the conclusion that, \emph{in the
  absence of mergers,} observed clusters with $M>\hmass$ should
contain on average $3.2$ massive galaxies (and certainly more if
accreted halos of lower masses also contain massive galaxies), with a
significant tail toward much larger numbers.  However, reproducing the
observed clustering of massive galaxies at $z\sim 0$ \citep{Zehavi05b}
would require closer to $1.2$ galaxies in such halos in our
simulation, in agreement with other work \citep{Masjedi06,
  Kulkarni07}.  While these statements are only qualitative, they will
be confirmed in the more quantitative discussion that follows.  In
order to reconcile the accretion properties of halos with
observations, we are thus lead to consider the fate of these massive
halos and the galaxies within them.

\section{The Merging Timescale of LRGs}\label{s:merge}

LRGs are massive galaxies with very little ongoing star-formation;
they thus constitute the tip of the red sequence.  They have uniform
spectral energy distributions marked by numerous features and hence
their redshifts are relatively straightforward to estimate
photometrically \citep[][redshift uncertainties are $\delta
z\sim0.03$]{Padmanabhan05}.  Modeling of their spectral energy
distributions has lead to the conclusion that these galaxies formed
the bulk of their stars at $z>2$ \citep[e.g.][]{Trager00, Jimenez06,
  Thomas05}, and hence are expected to evolve largely
dissipationlessly at $z<1$.  Their clustering strength is large,
suggesting that they live in massive dark matter halos $M>\amass$
\citep{Zehavi05b}.

Recently, \citet{Ho07} has measured the multiplicity function of LRGs
extracted from the Sloan Digital Sky Survey \citep[SDSS;][]{DR4} for
43 clusters over the redshift range $0.2<z<0.5$.\footnote{The primary
  spectral feature used to measure photometric redshifts of LRGs is
  the $4000\AA$ break; at $z<0.2$ this feature moves out of the SDSS
  bandpass filters.  Hence our sample is restricted to $z>0.2$.
  Although Ho et al.'s sample extends to $z\sim0.6$, for our purposes
  we truncate it at $z=0.5$ to limit the amount of possible evolution
  within the sample.} Cluster virial masses were derived from
\emph{ROSAT} $X$-ray data and range from
$10^{14.1}<M_{200}/M_\odot<10^{14.9}$.  The average stellar mass of
the LRGs in this sample is $M_{\rm star}=10^{11.6} M_\odot$, as
determined from a color-based stellar mass estimator \citep[for a
Chabrier IMF;][]{Bell03}.  These clusters contain on average $2.5$
LRGs.  The reader is referred to \citet{Ho07} for further details
regarding these observations.

In this sample there are approximately five clusters that contain no
LRGs\footnote{We say approximately because Ho et al. statistically
  remove interlopers based on photometric redshift uncertainties and
  hence clusters contain a non-integer number of LRGs.}.  When
plotting the observed multiplicity function below we include both the
one reported in Ho et al. and one where clusters with $N<1$ are
artificially assigned $N=1$.  This is done to afford a more robust
comparison to our simple model (see below) where we \emph{assume} that
each cluster halo contains at least one LRG at the center.  As
discussed below, our conclusions are insenstive to this distinction.

\begin{figure}
\begin{center}
\resizebox{3.5in}{!}{\includegraphics{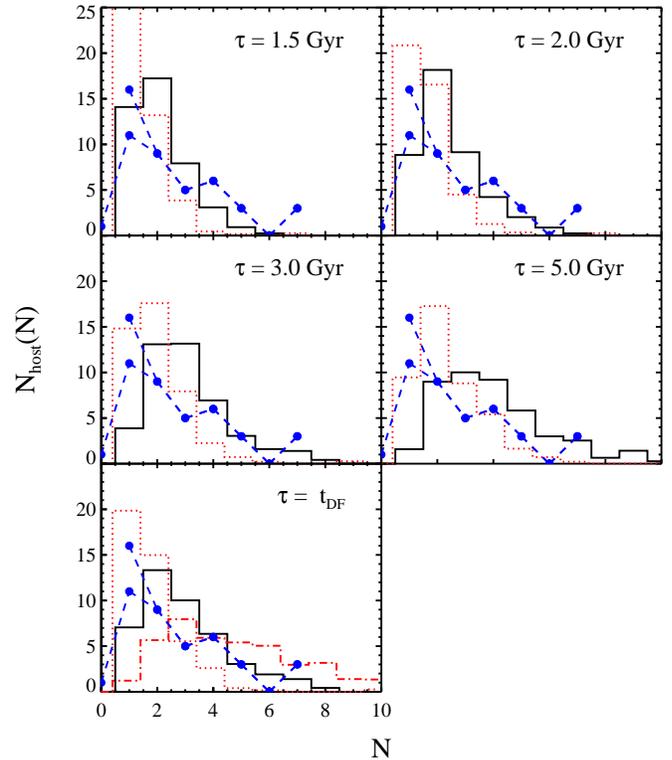}}
\end{center}
\caption{The multiplicity function of observed LRGs (\emph{dashed line
    joining points}) and accreted dark matter halos for two accreted
  mass thresholds: $M >\amass$ (\emph{solid line}) and $M >2\times
  10^{13} M_\odot$ (\emph{dotted line}).  The observed curves diverge
  at $N<1$ due to whether or not we artifically assign an LRG at the
  center of observed clusters with $N<1$ (see text for details).
  \emph{Top Four Panels:} Accreted halos are assumed to have disrupted
  after a time $\tau$, shown in the upper right corner of each panel.
  It is apparent that if LRGs can be identified with halos of mass
  $>\amass$ then they must on average merge within $\sim2$ Gyr.
  \emph{Bottom Panel:} Accreted halos are assumed to have merged after
  a dynamical friction timescale (see Equation \ref{eqn:df}).  This
  panel also includes halos more massive than $M>5\times10^{12}
  M_\odot$ (\emph{dot-dashed line}) for comparison.}
\label{fig:mf}
\end{figure}

The dark matter halo accretion history of massive halos (e.g. Figure
\ref{fig:mergers}) is closely related to the LRG multiplicity
function.  The former can be converted into the latter if one knows
both the minimum halo mass (measured at the epoch of accretion)
associated with accreted LRGs, $\mmin$, and the average time it takes
for LRGs to merge and/or disrupt\footnote{Throughout we use the words
  ``merge'' and ``disrupt'' interchangeably since our analysis does
  not distinguish between these two possibilities.} once accreted.
Below we argue for reasonable values of $\mmin$ and then attempt to
directly constrain the average LRG merging timescale.  We parameterize
the probability that an LRG will have merged by a time $t_{\rm{acc}}$
since accretion onto the host halo via:
\be
P_{\rm{merge}} = 1 - e^{-t_{\rm{acc}}/\tau}
\ee
where $\tau$ is the merging timescale.  The number of LRGs predicted
by this simple model is then 
\be
N_{\rm LRG} = 1 + \sum_i e^{-t_{\rm{acc,i}}/\tau}
\ee
where the first term counts one LRG at the center of the host halo and
the second term counts those satellites with accretion epoch mass
$>\mmin$ that have not merged.  For the purposes of generating a
multiplicity function we round the second term to the nearest integer.
Note that in generating a multiplicity function we do not have to make
a distinction between accretion events that did or did not occur
within the main host halo.  This distinction will only become relevant
when discussing the merger rates of LRGs.

\begin{figure}
\begin{center}
\resizebox{3.5in}{!}{\includegraphics{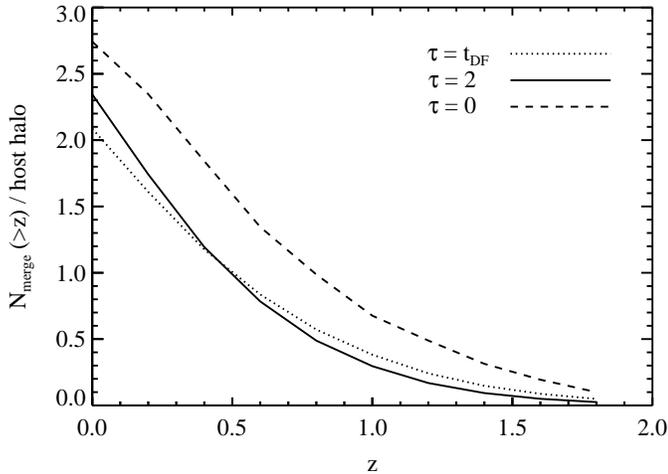}}
\end{center}
\caption{Number of accreted halos with $M>\amass$ that have merged
  with the host halo by redshift $z$, per $z=0$ host halo with
  $M>\hmass$.  The merging time is computed in two different ways: a
  constant timescale of either 0 or 2 Gyr (\emph{dashed} and
  \emph{solid lines}) and a timescale determined by the Chandrasekhar
  dynamical friction formula (\emph{dotted line}).}
\label{fig:mtime}
\end{figure}

The minimum LRG stellar mass in the observed sample is $10^{11.3}
M_\odot$.  A minimum halo mass associated with LRGs can be estimated
by assuming the universal baryon fraction $f_b=0.17$ and an efficiency
factor $\eta$ of converting baryons into stars.  While this factor is
only poorly constrained even at low redshift, values of order
$\sim0.1$ are likely reasonable for these massive galaxies
\citep{Hoekstra05,Mandelbaum06} and imply a minimum LRG hosting halo
mass of $\mmin\sim\amass$.  Note that the minimum halo mass for
hosting LRGs is here used to identify those accreted halos that are
likely to host an LRG, with a significant fraction of the accretion
occurring at $z\sim1$ or higher.  Thus this minimum mass will likely
not directly correspond to the minimum halo mass hosting LRGs at
$z\sim 0$, since the LRGs that have survived to the present epoch will
have accreted much more dark matter, resulting in a larger value for
$\mmin$ at the present epoch.  In our simulations halos today are on
average five times more massive than they were when they first crossed
$\mmin$, making our estimate consistent with that inferred from $z\sim
0$ clustering \citep{Zehavi05b,Kulkarni07}.  As we discuss in
\S\ref{s:stellar}, $\mmin$ much larger than $\amass$ would require
unreasonably long dynamical friction times and $\mmin$ either much
larger or much smaller would be in conflict with stellar mass
estimates in clusters.

Figure \ref{fig:mf} plots the resulting LRG multiplicity function both
for LRGs in accreted halos more massive than $\amass$ (solid lines)
and observations (dashed lines).  We also include predictions for LRGs
associated with halos twice as massive as our fiducial minimum mass
(dotted lines) in order to illustrate the sensitivity to our assumed
LRG halo mass threshold.  Each panel in Figure \ref{fig:mf} is the
multiplicity function for a different merger timescale.  The top four
panels assume that the merger timescale, $\tau$, is constant. It is
apparent from these panels that if $\mmin\sim\amass$ then LRGs must
merge on a characteristic timescale of $\sim2$ Gyr.  This timescale
implies an average number of LRGs per cluster of 2.5, satisfyingly
close to the observed value of 2.6.

Note that in order to compare to the observations we have weighted
host halos at $z=0.3$ (of which there are 460 in our simulation within
the observed mass range) in such a way as to reproduce the mass
distribution of the observed clusters.  It is the combination of these
two effects (higher redshift and different mass distribution) that
does not allow a direct comparison between Figures \ref{fig:mergers}
and \ref{fig:mf}.

If LRGs never merged, there would be on average $5.8$ LRGs per $z=0.3$
host halo (averaged over the observed distribution of halo masses).
Comparing this number to the observed $2.5$ LRGs per cluster
highlights the importance and prevalence of LRG mergers.  

The bottom panel in Figure \ref{fig:mf} assumes that the merger
timescale is equal to the dynamical friction timescale \citep{Binney87}:
\be
\label{eqn:df}
t_{\rm DF} = 0.1\,t_H\, \,\frac{M_h/M_s}{\ln(1+M_h/M_s)}
\ee
where $M_h$ and $M_s$ are the host and satellite masses, $z$ is the
redshift, $t_H$ is the Hubble time and all quantities are measured at
the epoch of accretion.  The average mass ratio at accretion is $\sim
6$ for our sample.  The pre-factor, $0.1\,t_H$, is the characteristic
time for a halo with mean density ${\mathcal O}(10^2)$ times the
critical density.  It is important to note that $t_{\rm DF}$ gets
shorter both at higher redshift and for merger mass ratios closer to
unity.  In this lower panel we have additionally included results for
halos of mass $M>5\times 10^{12} M_\odot$ (dot-dashed line) for
comparison.

This comparison with simple dynamical friction estimates provides a
satisfying cross check to the results in the upper panels.  In
particular, for our fiducial minimum LRG halo mass of $\amass$, the
dynamical friction timescale averaged over all the accreted halos is
$2.4\,$Gyr (median time is $1.8\,$Gyr --- the distribution is highly
asymmetric), which is quite similar to the constant merger timescale
that best matches the observed multiplicity function ($\sim2$ Gyr).
From Figure \ref{fig:mf} it is apparent that the simple DF timescale
would not have reproduced the observed LRG multiplicity function if
the minimum halos mass capable of hosting LRGs were substantially more
or less massive than $\amass$.  The implication here is clear: if
$\mmin$ is in fact considerably larger or smaller than $\amass$ then
simple DF arguments do not apply to the LRG population.

In fact, it is not at all clear that Equation \ref{eqn:df} should
apply here or in general to the dynamical evolution of satellite
galaxies, as it is strictly valid for a point mass moving in an
infinite, uniform background density field.  Indeed, much work has
gone into both testing the validity of Equation \ref{eqn:df} with
simulations \citep[e.g.][]{SWhite83,vdb99, Velazquez99,Read06} and
developing extensions to it \citep[e.g.][]{Tremain84,Colpi99},
including numerically following the evolution of the satellite orbit,
including the mass loss due to tidal forces \citep{Benson02,Taffoni03,
  Taylor04,Zentner05}.  This body of work has shown that Equation
\ref{eqn:df} is at best a crude approximation to the realistic,
time-dependent problem.  For these and other reasons it is quite
surprising, if not entirely coincidental, that the classical
Chandrasekhar DF timescale adequately captures the merging timescale
of LRGs.

As can be seen from the upper four panels in Figure \ref{fig:mf},
there is a degeneracy between the merger timescale and the minimum
halo mass associated with LRGs in the sense that a larger $\mmin$
coupled to a larger timescale can produce roughly the same
multiplicity function.  Thus, if one thought that LRGs lived in more
massive halos than what we have assumed here, then one would infer a
longer merging timescale for LRGs.  However, this is exactly opposite
to what one would infer from dynamical friction arguments since
$t_{\rm DF}\propto M_s^{-1}$.  Furthermore, increasing $\mmin$ to
$5\times\amass$ would result in far too few LRGs in massive halos
compared to observations, even if $\tau=\infty$; in this case the
average number of LRGs per halo would be $1.6$.  As we describe in
\S\ref{s:stellar}, $\mmin$ is further constrained by observations of
the stellar light in massive halos.

These merger timescales can easily be cast into a discussion of LRG
merger rates.  For this discussion we consider the full host halo
population at $z=0$, rather than the population at $z=0.3$ meant to
coincide with the data from Ho et al., in order to draw more general
conclusions about LRG mergers.  Figure \ref{fig:mtime} plots the
cumulative distribution of \emph{merged} LRGs as a function of
redshift, per host halo.  The figure includes constant merger
timescales of $0$ and $2\,$Gyr and a timescale set by dynamical
friction.  The $0\,$Gyr case can equivalently be thought of as the
distribution of \emph{accreted} LRGs, since in this case the accretion
and merging epochs are coincident.  In this figure we only count LRGs
that merge within the main progenitor of the $z=0$ host halo.  This
figure is thus not directly comparable to Figure \ref{fig:mergers}.
In other words, if a halo merges within a halo that itself later
mergers with the host halo then it is not counted here.  This plot is
thus meant to capture the number of mergers actually occurring
\emph{within the main progenitor} of the host halo.

The dynamical friction timescale is shorter than $2\,$Gyr at high
redshift and longer than $2\,$Gyr at low redshift; this results in a
more gradual increase in the merger rate per unit redshift compared to
a constant merger time of $2\,$Gyr.  Since the constant and dynamical
friction timescales are different at redshifts both greater and less
than $0.3$, comparisons to the multiplicity function at different
epochs can in principle rule out either (or both) of these timescales.
Both the constant $2\,$Gyr timescale and that determined by dynamical
friction imply that $2-2.5$ LRGs have merged with the host halo by
$z=0$.  Moreover, the figure indicates that a substantial number of
LRGs are merging/disrupting at $z<1$.  In the next section we set this
in the context of recent observational results of the stellar mass
budget in groups and clusters.

\begin{figure}
\begin{center}
\resizebox{3.5in}{!}{\includegraphics{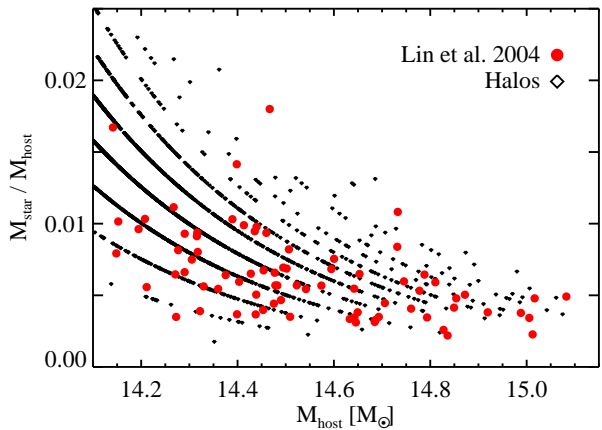}}
\end{center}
\caption{The total stellar-to-virial mass ratio as a function of $z=0$
  halo virial mass.  Only galaxies more massive than $M_{\rm
    star}=10^{11.3} M_\odot$ and the ICL are included in the stellar
  mass budget.  Observations from \citet[][\emph{solid
    circles}]{Lin04a} have been converted to total stellar masses, and
  are compared to total stellar masses estimated by assuming that
  every accreted halo with mass $>\amass$ hosts an LRG with $M_{\rm
    star}=10^{11.6} M_\odot$ (\emph{diamonds}).}
\label{fig:mstarmvir}
\end{figure}

\section{The Total Cluster Stellar Mass} \label{s:stellar}

The merger rate of LRGs found in the previous section indicates that a
significant number of LRGs must have disrupted over the history of
massive halos.  In particular, a merger timescale of $\tau=2\,$Gyr
implies that on average $2-3$ LRGs have disrupted within each host
halo more massive than $\hmass$ by $z=0$ (and approximately $2.1$
since $z=1$).  In this section we discuss how reasonable such a
disruption rate is in light of the total stellar mass observed in
clusters at $z\sim0$.

The amount of stellar mass brought into massive halos by LRGs can be
estimated in the following way.  We consider halos that have a mass at
$z=0$ greater than $\hmass$ and assign a stellar mass of $10^{11.6}
M_\odot$, the mean stellar mass of the observed LRG sample, to the
center of each.  Then, each halo that was accreted onto these $z=0$
halos is assigned the same amount of stellar mass if the halo mass at
accretion is $>\amass$.  This exercise is thus meant to count the
total amount of stellar mass that was at some point associated with
LRGs.  We make no distinction between disrupted and non-disrupted LRGs
except in one case: if, according to our best-guess LRG merger
timescale (2 Gyr), an LRG halo would have disrupted not in the main
progenitor of the $z=0$ halo but rather in some smaller halo that
would itself later accrete onto the main progenitor but does not
merge, then we do not count this LRG in the final stellar mass budget.
In this case the disrupted LRG contributes to the satellite's ICL
(i.e. ICL that surrounds the satellite and is distinct from the
central ICL).  As it turns out, only $20$\% of all accreted halos more
massive than $\amass$ fall into this category and including these
halos in the stellar mass budget does not appreciably change our
conclusions.

We compare to data presented in \citet{Lin04a} who have compiled
information on $93$ clusters at $z<0.1$, including $X$-ray
observations used to derive cluster virial masses and luminosities of
cluster members derived from 2MASS photometry.  From these data Lin et
al. have estimated the luminosity function (LF) of each cluster,
assuming that the faint-end slope is fixed at $\alpha=-1.1$.  Using
their LFs we are able to estimate the total luminosity in galaxies
brighter than $L_K=2.8\times10^{11} L_\odot$ which corresponds to the
minimum luminosity of the Ho et al. LRG sample.  This total luminosity
is converted into stellar mass by assuming a mass-to-light ratio of
$M_{\rm star}/L_K=0.72$ which is appropriate for red galaxies with a
Chabrier IMF \citep{Bell03}.  The LFs reported in \citet{Lin04a} do
not include the brightest cluster galaxy (BCG); we thus add these in
separately.  Finally, we have assumed that each cluster contains
intracluster light (ICL) with a stellar mass equal to the mass of the
BCG identified by Lin et al.~(i.e. $L_{\rm ICL} = L_{\rm BCG}$).  This
amount of light associated with the ICL is consistent with recent
observations \citep{Zibetti05, Gonzalez05}.  In this ``total'' cluster
luminosity we do not include any possible ICL associated with
satellites; it is for this reason that we did not include LRGs that
disrupted in halos which themselves later merged with the host.

Figure \ref{fig:mstarmvir} presents a comparison between the data from
\citet{Lin04a} and the stellar mass associated with accreted halos.
The agreement is encouraging.  Note that varying any one of our
assumptions can change the results from both the data and our model;
the important point to take away from this comparison is that the
influx of massive galaxies embedded within accreted halos appears to
roughly agree with the total stellar mass within observed clusters at
$z\sim0$.  This provides further support to our identification of
halos with mass $>\amass$ as being host to LRGs and suggests that
disrupted LRGs deposit their stars into a combination of the central
galaxy and ICL.  Increasing or decreasing $\mmin$ by a factor of two
would result in substantial disagreement with the observations shown
in Figure \ref{fig:mstarmvir}.  This is due to the fact that the
number of accreted halos does not scale linearly with the accreted
halo mass, and provides further support for our choice of
$\mmin=\amass$.

The simple model presented here also provides a straightforward means
for understanding the observed trend of decreased scatter in $M_{\rm
  star}/M_{\rm host}$ with increasing $M_{\rm host}$.  This arises
because the number of accreted halos with mass $>\amass$ is a weak
function of $M_{\rm host}$.  This is in contrast to the observed
number of satellites, which appears to be closer to linear in $M_{\rm
  host}$ \citep{Lin04a, Popesso07}.  The difference implies that
fractionally more halos/LRGs are merging in lower mass halos compared
to higher mass halos.

Interestingly, we find that it is not uncommon for $z=0$ host halos to
contain disrupted LRGs that did not disrupt within the host halo (and
hence were not counted in the above figure) but rather disrupted in a
smaller halo that later accreted onto the host (and remained as a
satellite to $z=0$).  If these disrupted LRGs are depositing some
fraction of their stars into ICL, then this suggests that there could
be a significant amount of ICL that is not centered on the central
galaxy but is instead centered on cluster satellites.  Such a scenario
is corroborated by recent observations of the Virgo cluster that show
significant amounts of ICL surrounding several of the most massive
satellites \citep{Mihos05}.

\section{Discussion}\label{s:conc}

The results of the previous sections suggest the following picture.
If LRGs are associated with halos more massive than $\amass$ at the
time when they are accreted onto more massive host halos, then the
observed multiplicity function of LRGs at $z\sim0.3$ implies that LRGs
must merge and/or disrupt on timescales of $\sim2$ Gyr.  Such a merger
rate implies that $2-3$ such LRGs have disrupted in halos more massive
than $\hmass$ by $z=0$.  This merger timescale is consistent with
classical dynamical friction arguments and suggests that a rather
simplistic dynamical prescription for the evolution of LRGs is
applicable when considering ensemble averages.

Moreover, the amount of total stellar mass in clusters that was at one
point associated with these infalling LRGs (ignoring for the moment
whether or not this stellar mass is locked up in satellite galaxies)
is consistent with observations when the observed amount of stars in
the intracluster light (ICL) is accounted for.  This in turn suggests
that the disrupting LRGs are depositing their stars into a combination
of the ICL and central galaxy, which is consistent with previous
modeling \citep{Monaco06,Murante07,Purcell07, Conroy07b}. Finally,
there appears to be a significant number of LRGs that have disrupted
within halos that only later accreted onto (but did not merge with)
what would become the $z=0$ host halo.  This suggests that there could
be a significant amount of ICL surrounding cluster satellites, in
addition to what is known to be associated with the central galaxy.

It has been historically challenging to constrain the merger rate of
galaxies.  Previous studies have relied on either morphological
disturbances \citep[e.g.][]{Conselice03, vanDokkum05, Bell06} or close
pair counts \citep[e.g.][]{Masjedi06} as probes of the merger rate of
massive galaxies.  Unfortunately, both methods are rather indirect
since the connection between either morphological disturbances or
close pair counts and merger rates is uncertain.  The most recent
inferred LRG-LRG merger rate is from \citet{Masjedi06} who find a rate
of $0.6\times 10^4$ Gyr$^{-1}$ Gpc$^{-3}$.  Averaging over all halos
between $z=0.5$ and $z=0.2$, the model presented herein implies an
LRG-LRG merger rate of $(1.0-1.3)\times 10^4$ Gyr$^{-1}$ Gpc$^{-3}$,
depending on whether the constant 2 Gyr or dynamical friction
timescale is used.  The agreement with \citet{Masjedi06} is
encouraging, especially given the (different) uncertainties in both
approaches.  These rates are also consistent with current predictions
from cosmological hydrodynamic simulations \citep{Maller06}.

Many studies have attempted to constrain the stellar mass growth of
massive galaxies from their inferred merger rates.  However, as argued
in \citet{Conroy07b} and herein, the merging of massive galaxies will
often not correspond to significant growth of the resulting galaxy
because a substantial amount of stars can be transfered to the ICL.
We hence caution against using merger rates to constrain the stellar
mass growth of galaxies.\footnote{This issue is intimately related to
  the way in which one counts galaxy light.  Of course the
  \emph{combined} light of both the central galaxy and its ICL will
  increase after a merger event.}  In fact, significant growth of the
ICL via merging at late times provides a means for reconciling two
apparently contradictory facts: one the one hand, observations at
$z<1$ indicate that central massive red galaxies grow little in mass
\citep[e.g.][]{Brown07, Fontana06, Bundy06, Wake06}, while on the
other hand, merging/disruption of galaxies within groups and clusters
at late times appears relatively common \citep[e.g.][]{MWhite07}.

\citet{MWhite07} outlined an approach for measuring the merging rate
of massive galaxies similar to the one presented herein.  Using the
observed evolution in the clustering of massive galaxies, these
authors concluded that $\sim1/3$ of massive satellites merge/disrupt
between $z\simeq0.9$ and $z\simeq0.5$.  In the present work we find
roughly $50$\% of massive satellites have disrupted over similar
epochs.  Our fraction is slightly higher because we have focused on
more massive galaxies than in \citet{MWhite07}.  The more general
conclusion from these two studies is, however, robust --- the
population of massive galaxies experiences significant amounts of
merging/disruption, even at $z<1$.

There is an important implication of considering the evolution of
galaxies within the context of the hierarchical growth of halos.  At
first glance, the lack of evolution in the observed correlation
function of massive galaxies and their halo occupation distribution at
$z<1$ suggests that massive galaxies do not disrupt or merge over this
epoch.  However, these galaxies are embedded within dark matter halos
that are continually merging and accreting new galaxies, which instead
suggests that massive galaxies \emph{must} merge in order that these
observed quantities not evolve appreciably at late times.  This
statement is further corroborated by dissipationless simulations which
show explicitly that the average number of subhalos within host halos
does not evolve appreciably at $z<1$ because the accretion and
disruption rate of subhalos are approximately equal \citep{Reed05,
  Conroy06a}.  If satellite galaxies reside within these subhalos then
the observed non-evolution of the clustering and halo occupation of
massive galaxies at $z<1$ is in fact consistent with significant
amounts of merging at late times.

Our results highlight the power of using purely dissipationless
simulations coupled to simple relations between galaxies and dark
matter to infer the evolution of galaxies and their relation to the
underlying dark matter with time.  The approach outlined herein can
easily be extended to other datasets to provide additional constraints
on the merger rate of galaxies.

\section*{Acknowledgments}

We thank David Hogg, Mike Boylan-Kolchin, Morad Masjedi, Eliot
Quataert, and Jerry Ostriker for comments on an earlier draft.
S.H. thanks David Spergel, Jim Gunn, Jerry Ostriker, Chris Hirata and
Nikhil Padmanabhan for insightful discussions and comments.
C.C. thanks nature for being complicated, but not too complicated.
M.W. is supported by NASA. The simulations were performed on the
supercomputers at the National Energy Research Scientific Computing
center.


\begin{appendix}
  
  \section{The Simulation, Halo Catalog, and Merger
    Trees} \label{sec:simulation}

We use a high resolution simulation of a $\Lambda$CDM cosmology
($\Omega_M=0.25=1-\Omega_\Lambda$, $\Omega_B=0.043$, $h=0.72$,
$n_s=0.97$ and $\sigma_8=0.8$).  The linear theory power spectrum is
computed by evolution of the coupled Einstein, fluid and Boltzmann
equations using the code described in \citet{MWhite96}.  This code agrees
well with {\sl CMBfast\/} \citep{Seljak96}, see e.g.~\citet{Seljak03}.  The
simulation employs $1024^3$ particles of mass $8\times
10^{9}\,h^{-1}M_\odot$ in a periodic cube of side $500\,h^{-1}$Mpc
using a {\sl TreePM\/} code \citep{MWhite02}.  The Plummer equivalent
softening is $18\,h^{-1}$kpc (comoving).

The phase space data for the particles exists at 50 outputs, spaced
equally in conformal time between $z\simeq 3$ and $z=0$.  For each
output we generate a catalog of halos using the Friends-of-Friends
(FoF) algorithm \citep{Davis85} with a linking length of $0.168\times$
the mean inter-particle spacing.  This procedure partitions the
particles into equivalence classes, by linking together all particle
pairs separated by less than a distance $b$.  The halos correspond
roughly to particles with $\rho>3/(2\pi b^3)\simeq 100$ times the
background density.  For each halo we compute a number of properties,
including the mass $M_{200}$ interior to $r_{200}$ within which the
mean density is $200\times$ the critical density.  $M_{200}$ is
computed from a fit of an NFW profile \citep{NFW97} to the particles
in the FoF group.

Merger trees are computed from the set of halo catalogs by identifying
for each halo a ``child'' at a later time.  The child is defined as
that halo which contains, at the later time step, more than half of
the particles in the parent halo at the earlier time step (weighting
each particle equally).  For the purposes of tracking halos this
simple linkage between outputs suffices (note that we do not attempt
to track subhalos within larger halos, which generally requires
greater sophistication).  From the merger trees it is straightforward
to compute the time when a halo `falls in' to a larger halo, the
number and masses of the progenitors etc.

\end{appendix}

\end{document}